\pdfoutput=1

\documentclass[aps,prl,twocolumn,nofootinbib,superscriptaddress,preprintnumbers]{revtex4-1}  

\usepackage{graphicx}  
\usepackage{amsmath,amssymb}  
\usepackage{hyperref}
\usepackage{color}
\graphicspath{{./Figures/}}
\usepackage[utf8]{inputenc}

\hypersetup{
    colorlinks=true,    
    linkcolor=red,      
    citecolor=blue,     
    filecolor=magenta,  
    urlcolor=blue       
}

\begin{document}

\title{Tensor Modes in Pure Natural Inflation}

\author{Yasunori Nomura}
\affiliation{Berkeley Center for Theoretical Physics, Department of Physics,
 University of California, Berkeley, CA 94720, USA}
\affiliation{Theoretical Physics Group, Lawrence Berkeley National Laboratory,
 Berkeley, CA 94720, USA}
\affiliation{Kavli Institute for the Physics and Mathematics of the 
 Universe (WPI), University of Tokyo, 
 Kashiwa, Chiba 277-8583, Japan}

\author{Masahito Yamazaki}
\affiliation{Kavli Institute for the Physics and Mathematics of the 
 Universe (WPI), University of Tokyo, 
 Kashiwa, Chiba 277-8583, Japan}
 
\preprint{IPMU17-0168}

\begin{abstract}
We study tensor modes in pure natural inflation~\cite{Nomura:2017ehb}, 
a recently-proposed inflationary model in which an axionic inflaton couples 
to pure Yang-Mills gauge fields.  We find that the tensor-to-scalar ratio 
$r$ is naturally bounded from below.  This bound originates from the 
finiteness of the number of metastable branches of vacua in pure Yang-Mills 
theories.  Details of the model can be probed by future cosmic microwave 
background experiments and improved lattice gauge theory calculations 
of the $\theta$-angle dependence of the vacuum energy.
\end{abstract}

\maketitle


Cosmic inflation is a successful framework in explaining many distinguished 
features of our Universe, including its flatness and the origin of 
primordial density perturbations.  There are, however, a plethora of 
inflationary models proposed in the literature, and we ultimately need 
to turn to observations for guidance, to convincingly answer the question 
of exactly which inflationary model describes our Universe.

Future detection of primordial tensor modes in comic microwave background 
(CMB) radiation would be ideal for this purpose.  The size of primordial 
tensor modes is quantified by the tensor-to-scalar ratio $r$, and when 
combined with the observed value of the scalar spectral index $n_s$, 
these two parameters severely constrain models of inflation.  This 
therefore provides an exciting opportunity for narrowing down possible 
models, especially because values of $r \sim 10^{-3}$ are expected 
to be within reach in next-generation CMB measurements (see e.g.\ 
Ref.~\cite{Creminelli:2015oda}).

The goal of this paper is to study the prediction for tensor modes 
of the recently-proposed inflationary model of {\it pure natural 
inflation}~\cite{Nomura:2017ehb}.  This is arguably the simplest model 
of inflation consistent with the current observational data.  It is 
defined within conventional low-energy effective field theory and is 
technically natural, i.e.\ stable under quantum corrections.

The model is given by an axionic inflaton $\phi$ coupling to 
four-dimensional pure Yang-Mills gauge fields:
\begin{equation}
  \mathcal{L}_{\phi F F} = \frac{1}{32\pi^2} \frac{\phi}{f} 
    \epsilon^{\mu\nu\rho\sigma} \textrm{Tr} F_{\mu \nu} F_{\rho \sigma},
\label{eq:phiFF}
\end{equation}
where $f$ is the decay constant and the dimensionless combination 
$\theta:=\phi/f$ plays the role of the $\theta$-angle of the Yang-Mills 
theory.  Below we choose the Yang-Mills gauge group to be $SU(N)$ 
for simplicity.

The inflaton potential $V(\phi)$ is determined by the dynamics of the 
pure Yang-Mills theory.  For our purposes, it is useful to parameterize 
the potential in the form
\begin{equation}
  V(\phi) = M^4 \left[ 1 - \frac{1}{\left( 1 + (\phi/F)^2 \right)^p} \right].
\label{eq:V-gen}
\end{equation}
Here, $M$ and $F$ are two parameters which have dimensions of mass, and 
the exponent $p > 0$ is a dimensionless parameter.  The parameter $F$ 
plays the role of the effective decay constant.

This potential is motivated by the holographic computation of 
Ref.~\cite{Dubovsky:2011tu}, which gives the parameters $M$ and $F$ 
to be
\begin{equation}
  M \approx \sqrt{N} \Lambda,
\qquad
  F \approx N f,
\label{eq:MFp}
\end{equation}
where $\Lambda$ is the dynamical scale of the Yang-Mills theory.  We 
define the parameter $\gamma$ by
\begin{equation}
  F = \pi \gamma N f.
\label{eq:gamma}
\end{equation}
As we will see later, $\gamma \approx O(1)$.  For our purposes, we use 
$\gamma$ and the power $p$ to parameterize the strong-coupling dynamics 
of the Yang-Mills theory.%
\footnote{The holographic result in Ref.~\cite{Dubovsky:2011tu} gives 
 $p = 3$.  We will not be restricted to this specific value; see 
 Ref.~\cite{Nomura:2017ehb}.}

The potential of Eq.~(\ref{eq:V-gen}) takes approximately the quadratic 
form $V(\phi) \sim \phi^2$ for $\phi \ll F$, but it begins to deviate 
from this form as $\phi/F$ becomes larger.  Note that this potential 
is rather different from the cosine potential used in the original model 
of natural inflation~\cite{Freese:1990rb,Adams:1992bn}, which is motivated 
by the instanton approximation---as explained in Ref.~\cite{Nomura:2017ehb}, 
the cosine potential is not theoretically valid for pure Yang-Mills 
theory, and is also disfavored by the recent observations by 
Planck~\cite{Ade:2015lrj} and BICEP2/Keck Array~\cite{Array:2015xqh}.

\begin{figure}[t]
\includegraphics[scale=0.67]{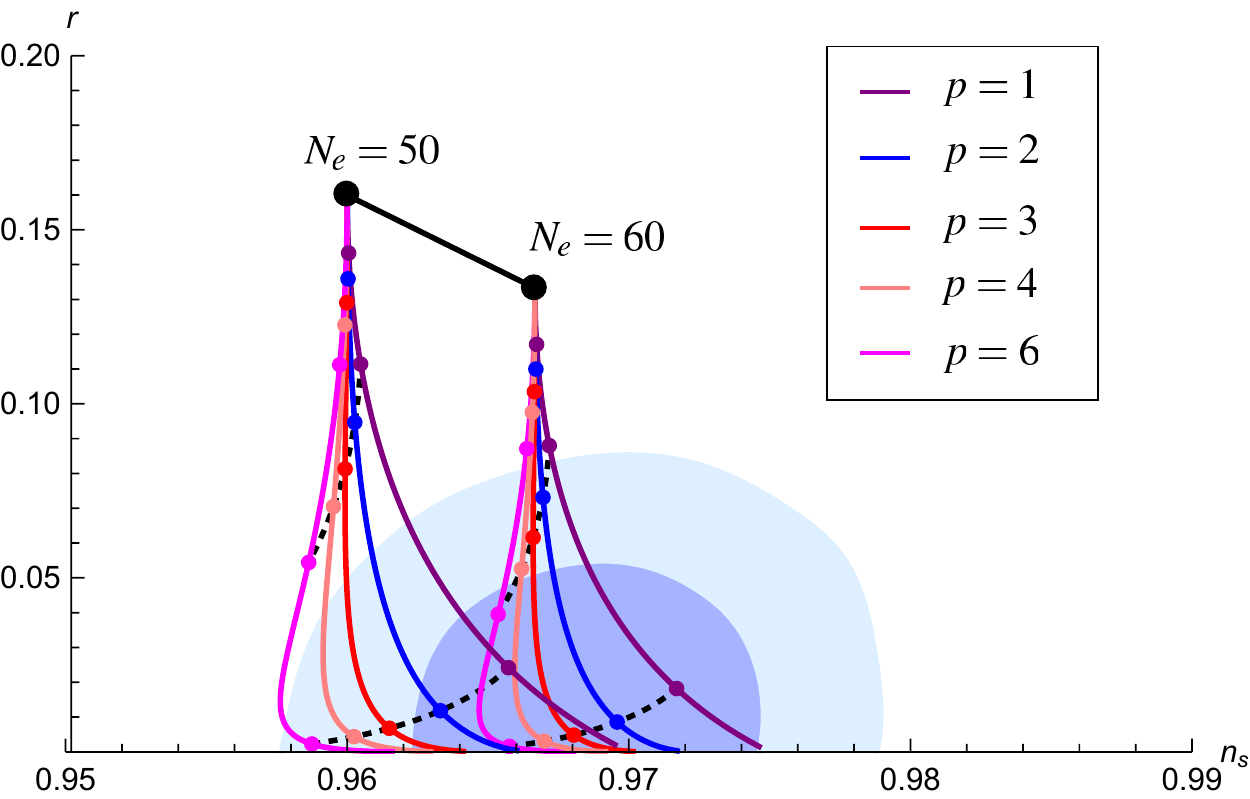}
\caption{The values of $n_s$ and $r$ predicted by the model for the 
 number of e-folds $N_e = 50, 60$ and for $p = 1,2,3,4,6$.  The light 
 (dark) blue region represents the 95\% (68 \%) CL allowed region by 
 Planck~\cite{Ade:2015lrj} and BICEP2/Keck Array~\cite{Array:2015xqh}. 
 $F/M_{\rm Pl} = 10,5,1$ are indicated by dots (from top to bottom). 
 This plot is the same as that in Fig.~3 of Ref.~\cite{Nomura:2017ehb}, 
 except for the choice of the values of $p$.}
\label{fig:rns}
\end{figure}

The parameter $M$ in Eq.~(\ref{eq:V-gen}) is determined by the overall 
size of the scalar perturbation once the other parameters, $F$ and $p$, 
are given.  While the power spectrum depends on all these parameters, 
in the range of $F$ and $p$ considered in this paper, we find
\begin{equation}
  M \sim 10^{16}~{\rm GeV}.
\label{eq:M}
\end{equation}
This implies that to discuss the tensor-to-scalar ratio $r$ and spectral 
index $n_s$, only the effective decay constant $F$ and the power $p$ are 
relevant.  When we vary these parameters, we obtain a range of $r$ and 
$n_s$ which are in impressive agreement with the current observational 
constraints; see Fig.~\ref{fig:rns}.

We see that the value of the spectral index $n_s$ is mostly consistent 
with observation regardless of the values of $F$ and $p$.  On the other 
hand, the size of the tensor-to-scalar ratio $r$ strongly depends on 
the value of $F$.  Our main interest in this paper is to figure out the 
expected size of the tensor-to-scalar ratio $r$, or equivalently the 
value of $F$, in the present model.

If $F$ is large, $F \gtrsim M_{\rm Pl}$, we expect to have a large 
value of $r$, and hence tensor modes can be observationally found in 
the near future.  Here, $M_{\rm Pl} \simeq 1.22 \times 10^{19}~{\rm GeV}$ 
is the Planck scale.  In the limit that $F$ is very large, 
$F \gg M_{\rm Pl}$, the prediction of the model approaches that of 
chaotic inflation~\cite{Linde:1983gd} with the quadratic potential 
$V(\phi) = m^2 \phi^2/2$, which is now excluded at about a $3\sigma$ 
level~\cite{Ade:2015lrj,Array:2015xqh}.  However, as discussed in 
our previous paper~\cite{Nomura:2017ehb}, this extreme limit is not 
available in our framework, since the validity of low-energy effective 
field theory puts a constraint $F \lesssim O(M_{\rm Pl})$.

In the opposite limit of small $F$, the tensor-to-scalar ratio $r$ is 
small; in fact, it can be tiny if $F$ is much smaller than $M_{\rm Pl}$. 
At first sight, there seems to be no issue in going to this extreme 
parameter region.  The predicted value of $n_s$ is consistent with 
current experimental bounds, as can be seen in Fig.~\ref{fig:rns}. 
The necessary amount of inflation, $N_e \approx 50\mbox{--}60$, can 
be obtained if the initial value of the inflaton field is large, 
$\phi \gg F$.  However, there is a reason to think that such a parameter 
region may not be available in the model.  This has to do with the 
fact that the potential in Eq.~(\ref{eq:V-gen}) is motivated by the 
holographic computation in the large $N$ limit, and it should not be 
taken at face value once we taken into account the finite $N$ effects.

To explain this point (in the language of quantum field theory), 
let us first recall the salient features of the large $N$ 
analysis~\cite{Witten:1979vv,Witten:1980sp}.

In addition to the axion coupling in Eq.~(\ref{eq:phiFF}), we have the 
kinetic term for the gauge fields, so that the total Lagrangian density 
is given by
\begin{equation}
  \mathcal{L} 
  = N \left[ -\frac{1}{4\lambda} \textrm{Tr}(F^{\mu \nu} F_{\mu\nu}) 
    + \frac{1}{32\pi^2} \frac{\phi}{N f} \epsilon^{\mu\nu\rho \sigma} 
    \textrm{Tr} F_{\mu \nu} F_{\rho \sigma} \right].
\label{eq:L_total}
\end{equation}
Here, we have factored out the overall coefficient $N$, and $\lambda = 
g^2 N$ is the 't~Hooft coupling with $g$ being the gauge coupling.  In 
the large $N$ limit~\cite{tHooft:1973alw}, the parameter $1/N$ plays the 
role of an expansion parameter.  Physical observables are expected to 
be smooth functions of $\lambda$ and $\phi/(N f)$, which are kept finite 
in taking the limit.

From this large $N$ scaling argument, we expect that the potential of 
$\phi$, i.e.\ the $\theta$-angle dependence of the vacuum energy, takes 
the form
\begin{equation}
  V(\phi) = N^2 \Lambda^4 \mathsf{V}\left(\frac{\phi}{N f} \right) + O(N^0),
\label{eq:V_N}
\end{equation}
where $\mathsf{V}(x)$ is a smooth function of $O(N^0)$ when written in 
terms of $x$.  This potential, however, does not respect the expected 
symmetry under $\phi \to \phi + 2\pi f$.  In the large $N$ limit, this 
transformation induces an infinitesimal shift in the argument of function 
$\mathsf{V}$, which can be an invariance of the potential $V(\phi)$ 
only if $\mathsf{V}$ is constant.  However, this is inconsistent with 
perturbative large $N$ calculation, which shows otherwise.

The way around this problem is to realize that the potential is 
multi-valued~\cite{Witten:1980sp}.  In particular, we have many different 
(in general metastable) branches corresponding to the shift $\phi \to 
2\pi f n$ with $n$ integer.  The correct vacuum energy, for example, 
is then given by the minimal values among these branches
\begin{equation}
  V_{\rm min}(\phi) = N^2 \Lambda^4 
    \textrm{min}_n \mathsf{V}\left(\frac{\phi+2\pi f n}{N f} \right),
\label{eq:V_min}
\end{equation}
so that the invariance of physics under $\phi \to \phi + 2\pi f$ is 
recovered.

Let us now come to finite values of $N$.  In the large $N$ analysis the 
value of $N$ is taken to be infinity, so that we have an infinitely many 
branches, i.e.\ $n$ runs over all integers in Eq.~(\ref{eq:V_min}).%
\footnote{See, e.g., Refs.~\cite{Kaloper:2011jz,Dubovsky:2011tu,%
 Dine:2014hwa,Yonekura:2014oja,Kaloper:2016fbr,DAmico:2017cda} for 
 related discussion in the context of inflation.}
However, the situation can be different for a finite value of $N$---if 
$n$ is taken to be of order $N$ then the shift $\phi \to \phi + 2\pi f n$ 
changes the argument of $\mathsf{V}(\phi/(N f))$ by an $O(1)$ amount, which 
can preserve the value of the function $\mathsf{V}(\phi/(N f))$.  If this 
happens, there will be only a finite number of metastable branches, with 
each branch being periodic with the period of $O(2\pi N f)$.

That a finite number (order $O(N)$) of branches exists is discussed 
in the analysis of the chiral Lagrangian for QCD (with flavors) in 
Ref.~\cite{Witten:1980sp}.  The analysis there is justified for small 
quark masses, whereas here we are interested in the opposite limit 
of pure Yang-Mills theory, in which the quark masses are taken to be 
infinitely large.  

In the case of pure Yang-Mills theory, we expect that the number 
of metastable branches is $N$ (so that the periodicity of the 
$\theta$-dependent potential in a single branch is $2\pi N$, not $2\pi$). 
This is suggested for example by the analysis of softly-broken 
$\mathcal{N} = 1$ supersymmetric Yang-Mills theories (see 
Refs.~\cite{Dine:2014hwa,Yonekura:2014oja}).  More recently, this 
$2\pi N$ periodicity of the $\theta$ angle has been made manifest 
when we turn on the background gauge fields for the center of the 
gauge group~\cite{Gaiotto:2017yup}.

Yet another support comes from the recent work of 
Ref.~\cite{Yamazaki:2017ulc} (see also~\cite{Yamazaki:2017dra}), 
which has given a concrete argument that the number of branches 
is $N$ (so that we have $\mathsf{V}(x+2\pi) = \mathsf{V}(x)$). 
This paper studied a compactification of pure $SU(N)$ Yang-Mills 
theory on $\mathbb{R} \times \mathbb{T}^3$, twisted by the $\mathbb{Z}_N$ 
center symmetry.  There the $\theta$ angle of the Yang-Mills theory 
is identified with the $\theta$ angle of the $\mathbb{CP}^{N-1}$ model, 
where $\mathbb{CP}^{N-1}$ arises as the moduli space of flat connections 
on $\mathbb{T}^2$.  The classical vacua of the theory are given 
by the $N$ fixed points of the twisted boundary condition of the 
$\mathbb{CP}^{N-1}$ model, thereby identifying the $N$ vacua explicitly 
in a weakly coupled region.

\begin{figure}[htbp]
  \centering\includegraphics[scale=0.8]{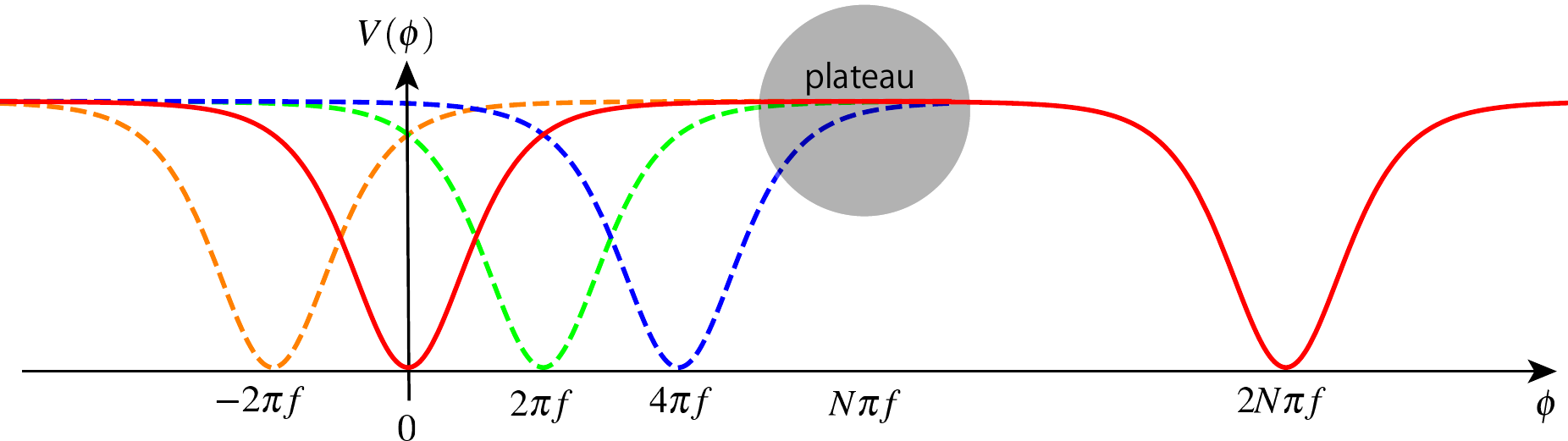}
\caption{A sketch of the inflaton potential $V(\phi)$ for a finite value 
 of the gauge group rank $N$.  We show the branch structure corresponding 
 to the shift of $\phi$ by $2\pi n f$ ($n = 1,2,\ldots$) with different 
 colors.  Each metastable branch has a period of $2\pi N f$ to account 
 for the $N$-fold degeneracy of metastable vacua.  We have a plateau 
 near the value $\phi \sim \pi N F$ (for a branch that has a minimum 
 at $\phi = 0$), which however has only a finite width for consistency 
 with the $2\pi N f$ periodicity.}
\label{fig:pot_finiteN}
\end{figure}

Given all the evidence, we assume below that the number of different 
branches is $N$:
\begin{equation}
  V(\phi + 2\pi N f) = V(\phi).
\label{eq:V-periodic}
\end{equation}
In addition, we have parity symmetry $V(\phi) = V(-\phi)$ and know 
the form of the potential near the origin, $\mathsf{V}(x) \sim x^2$ 
from perturbative large $N$ calculation.  These imply that the potential 
should have a plateau region of finite size around $\phi \sim \pi N f$; 
see Fig.~\ref{fig:pot_finiteN}.  Here and below, we focus on the branch 
that has a minimum at $\phi = 0$ without loss of generality.

\onecolumngrid

\twocolumngrid

An important point is that the plateau portion of the potential, which 
is useful for inflation, cannot continue indefinitely, since the potential 
decreases as $\phi$ deviates far from a minimum, e.g.\ in the region 
$N \pi f < |\phi| < 2N \pi f$ when viewed from $\phi = 0$.  This implies 
that the value of the inflaton corresponding to the current horizon scale, 
which we denote by $\phi_*$, can be restricted to the region $\phi_* \in 
[0, N\pi f]$ without loss of generality.  (If the initial value of the 
inflaton $\phi_i$ is larger than $N\pi f$, then the inflaton will roll 
into another minimal $2N\pi f$, resulting in the identical physics as 
the inflaton starting from $2\pi f - \phi_i$.)  Note that we do not know 
the precise form of the potential around the turning point $\phi \sim 
N\pi f$.  However, the potential of Eq.~(\ref{eq:V-gen}) is already quite 
flat around these points, and we expect that it smoothly connects to the 
$2N \pi$ shifted potential, as depicted in Fig.~\ref{fig:pot_finiteN}.

Let us define the variable
\begin{equation}
 y \equiv \frac{\phi_*}{F}.
\label{eq:phi_limit}
\end{equation}
Since $y \approx \phi_*/N\pi f$ (see Eq.~(\ref{eq:gamma})), the parameter 
space for this quantity is bounded
\begin{equation}
  y \leq \frac{1}{\gamma} \equiv y_{\rm max} \approx O(1),
\label{eq:y_max}
\end{equation}
where $y = y_{\rm max}$ corresponds to $\phi_* = N\pi f$; see 
Fig.~\ref{fig:pot_finiteN}.  In this paper, we do not consider the 
case in which $y$ is extremely close to $y_{\rm max}$, i.e.\ the case 
in which observable inflation starts near $\phi = N\pi f$.

There are several reasons for this.  First, for $\phi$ very close to 
$N\pi f$, we cannot trust the form of the potential in Eq.~(\ref{eq:V-gen}). 
In this region, the potential becomes flat, with $dV(\phi)/d\phi = 0$ 
at $\phi = N\pi f$.  Since the potentials of many branches are almost 
degenerate there, we suspect that tunnelings between different branches 
may not be negligible.%
\footnote{It may be interesting to consider a scenario in which 
 $y \approx y_{\rm max}$, if these tunnelings are small.  This could 
 lead to small field inflation with the value of $F$ being very small, 
 resulting in negligible $r$.  We do not pursue this possibility 
 further.}
Furthermore, one might argue that natural values for $y$ are not close 
to $y_{\rm max}$.  Since the axion is a pseudo Nambu-Goldstone boson, 
the inflaton potential is regarded as a flat direction at energy scales 
much larger than $M$.  One may then expect that the initial value 
of the inflaton $\phi_i$ is distributed uniformly in the interval 
$[0, N \pi f]$.  If this is the case, then we expect
\begin{equation}
  y \lesssim 0.5 y_{\rm max},
\label{eq:y_ave}
\end{equation}
``on average.''  Below, we consider that
\begin{equation}
  y \lesssim 0.9 y_{\rm max}.
\label{eq:y_bound}
\end{equation}
We assume that in this parameter region, the potential of 
Eq.~(\ref{eq:V-gen}) can be used to predict $n_s$ and $r$.

The prediction for $r$ depends somewhat sensitively on $y_{\rm max}$, 
i.e.\ the value of $\gamma$ in Eq.~(\ref{eq:gamma}).  This parameter, 
however, can be determined from first-principle computations in lattice 
gauge theory.

To see this, let us expand the potential in power series in 
$\theta = \phi/f$:
\begin{equation}
  V(\theta) = \frac{1}{2} \chi \theta^2 
    \left( 1 + \sum_{n=1}^{\infty} b_{2n} \theta^{2n} \right),
\label{eq:V-theta}
\end{equation}
where the leading coefficient $\chi$ is known as the topological 
susceptibility, and the sub-leading coefficients have the large $N$ 
expansion~\cite{Witten:1980sp}
\begin{equation}
  b_{2n} = \frac{\bar{b}_{2n}}{N^{2n}} 
    \left( 1 + O\left(\frac{1}{N^2}\right) \right).
\label{eq:b2n_N}
\end{equation}
The potential of Eq.~(\ref{eq:V-gen}) gives values%
\footnote{We have $\chi = (2p M^4/\pi^2 \gamma^2 N^2)(1 + O(1/N^2)) 
 \approx O(p \Lambda^4)$ for the topological susceptibility.  Note that 
 our potential, Eq.~(\ref{eq:V-gen}), keeps only the leading terms in 
 the large $N$ expansion, Eq.~(\ref{eq:b2n_N}), of the coefficients 
 $b_{2n}$.}
\begin{equation}
  \bar{b}_2 = -\frac{p+1}{2} \left(\frac{1}{\pi\gamma} \right)^2,
\quad
  \bar{b}_4 = \frac{(p+1)(p+2)}{6} \left(\frac{1}{\pi\gamma} \right)^4.
\label{eq:b2-b4_1}
\end{equation}
This should be compared with the recent results from lattice gauge 
theory~\cite{Bonati:2016tvi}
\begin{equation}
  \bar{b}_2 = -0.23(3),
\quad
  \bar{b}_4 \lesssim 0.1.
\label{eq:b2-b4_2}
\end{equation}
From this, $\gamma$ is determined in terms of $p$ as
\begin{equation}
  \gamma \simeq 0.47 \sqrt{p+1},
\label{eq:gamma-latt}
\end{equation}
giving rise to the constraint
\begin{equation}
  y \lesssim 0.9 \frac{2.1}{\sqrt{p+1}} 
    \simeq \frac{1.9}{\sqrt{p+1}}.
\label{eq:y_upper}
\end{equation}

\begin{figure}[t]
\includegraphics[scale=0.6]{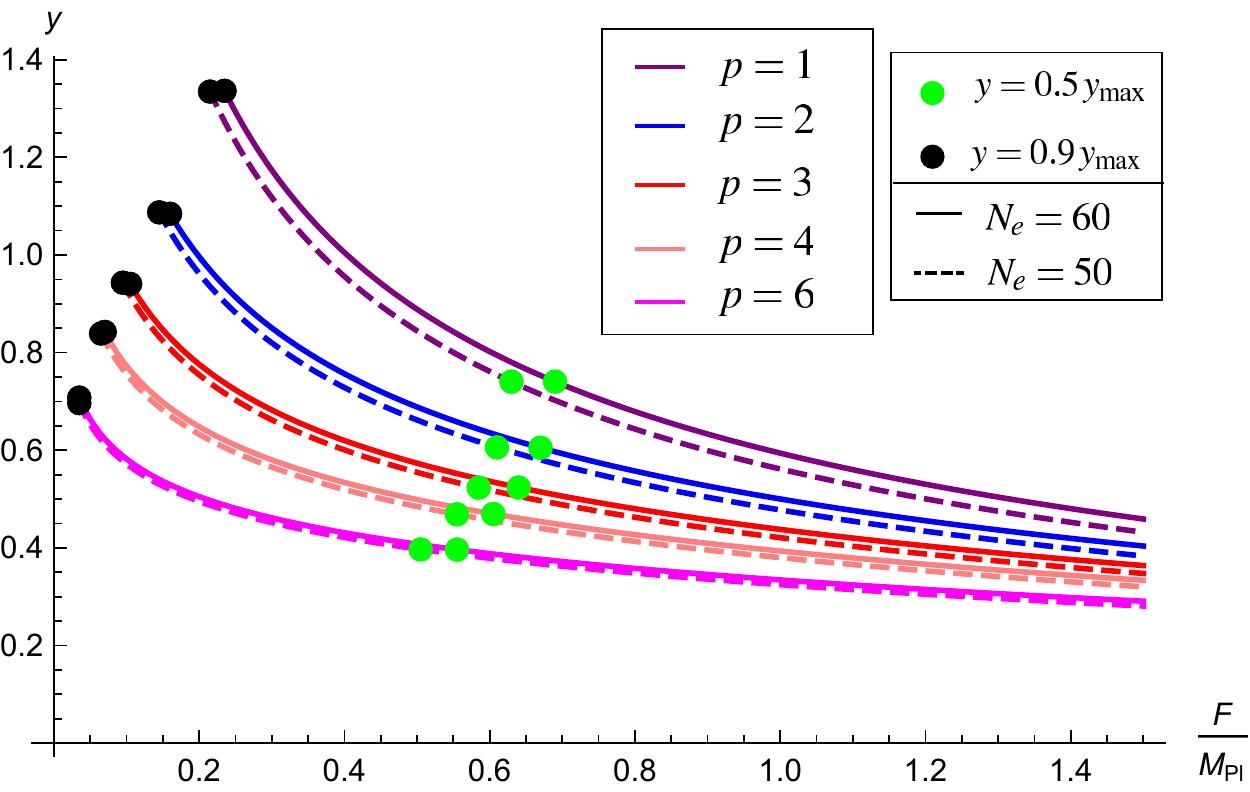}
\caption{The value of $y = \phi_*/F$ as a function of $F/M_{\rm Pl}$ 
 for $N_e = 60$ (solid) and $50$ (dashed) for $p=1,2,3,4,6$ (from 
 top to bottom).  Here, $\phi_*$ is the value of the inflaton at 
 which observable inflation starts (i.e.\ inflation occurring 
 in the region $\phi \leq \phi_*$ has $N_e$ e-folds).  Black dots 
 represent $y = 0.9 y_{\rm max}$ with $y_{\rm max} = 2.1$; see 
 Eq.~(\ref{eq:y_upper}).  Green dots represent $y = 0.5 y_{\rm max}$.}
\label{fig:F-y}
\end{figure}

\begin{figure}[t]
\includegraphics[scale=0.6]{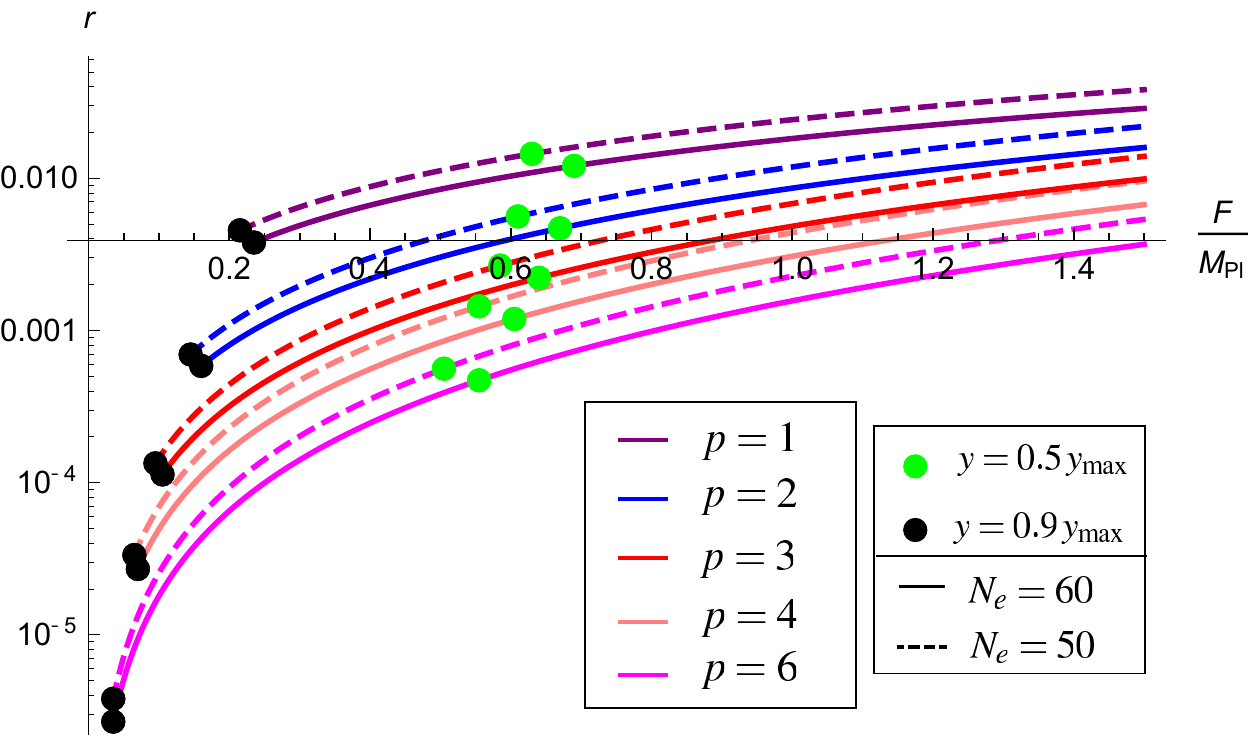}
\caption{The tensor-to-scalar ratio $r$ as a function of $F/M_{\rm Pl}$ 
 for $N_e = 60, 50$ for $p=1,2,3,4,6$.  Lines and dots are the same as 
 in Fig.~\ref{fig:F-y}.}
\label{fig:F-r}
\end{figure}

This constraint can be translated into a lower bound on the effective 
decay constant $F$, and then into a lower bound on the tensor-to-scalar 
ratio $r$.  In Fig.~\ref{fig:F-y}, we plot the value of $y$ as a function 
of $F/M_{\rm Pl}$ for $p = 1,2,3,4, 6$, assuming that the number of e-folds 
for observation inflation (inflation occurring in $\phi \leq \phi_*$) 
is $N_e = 50, 60$.  The black and green dots correspond to $y = 0.9 
y_{\rm max}$ and $0.5 y_{\rm max}$, respectively.  We find that with 
$y < 0.9 y_{\rm max}$, $F$ is bounded from below.  In Fig.~\ref{fig:F-r}, 
we give a similar plot, but now the vertical axis is $r$.  We find that 
the value of $r$ is bounded from below.
\begin{figure}[t]
\includegraphics[scale=0.6]{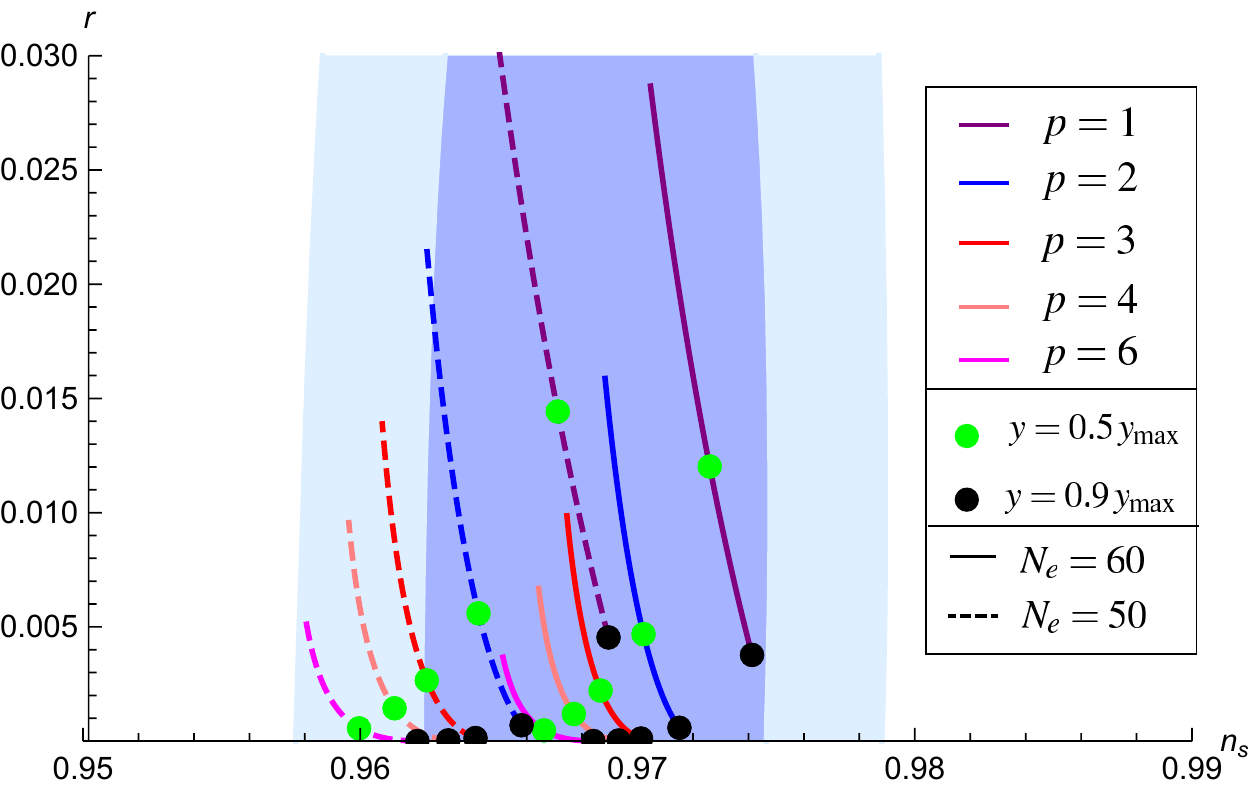}
\caption{Predictions of the model on the $n_s$-$r$ plane.  Lines and 
 dots are the same as in Fig.~\ref{fig:F-y}; in particular, black (green) 
 dots represent the bound $y < 0.9 y_{\rm max}$ ($0.5 y_{\rm max}$).  The 
 light (dark) blue region represents the allowed parameter region under 
 the 95\% (68\%) CL from Planck~\cite{Ade:2015lrj} and BICEP2/Keck 
 Array~\cite{Array:2015xqh}.}
\label{fig:final-1}
\end{figure}
\begin{figure}[t]
\includegraphics[scale=0.6]{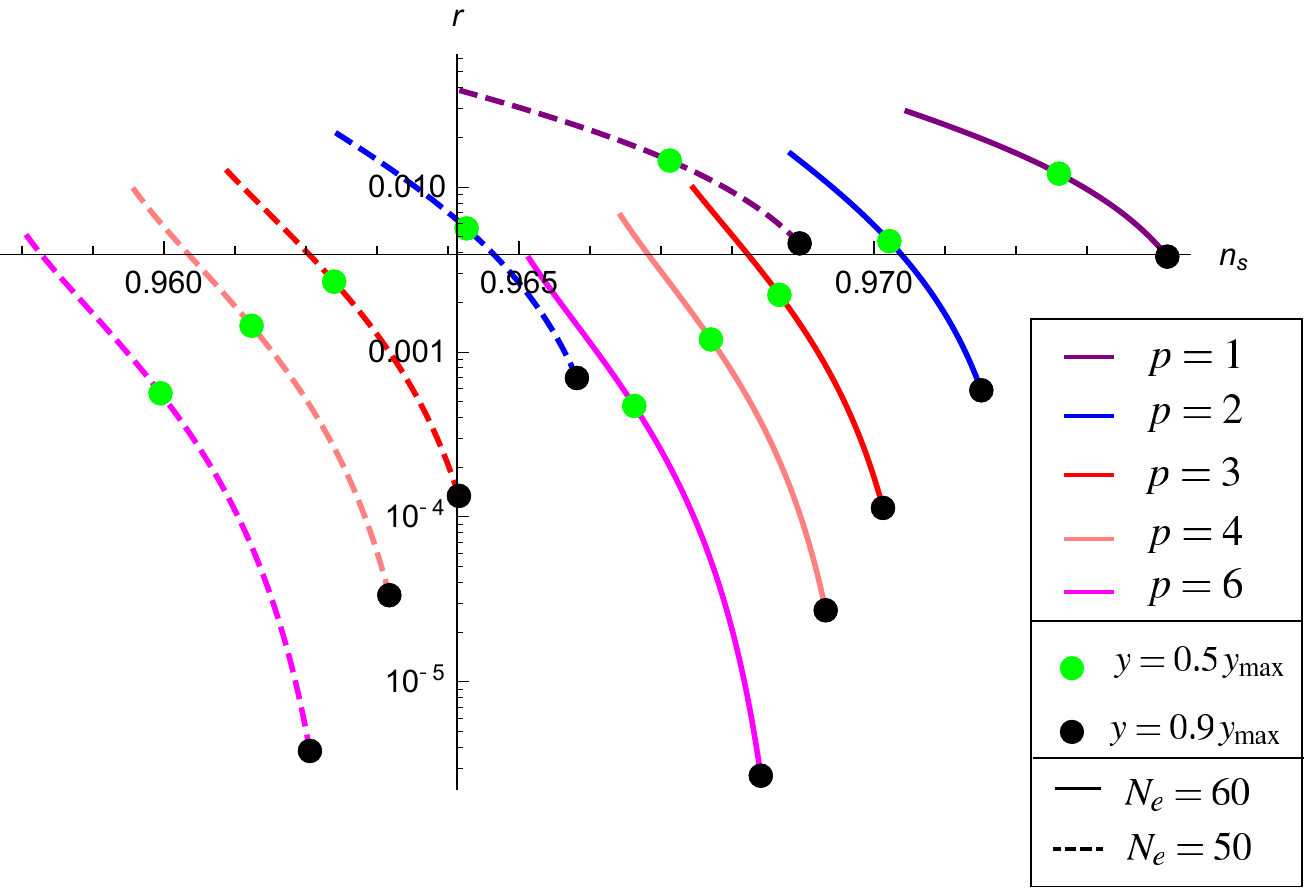}
\caption{The same as in Fig.~\ref{fig:final-1} but zoomed into the region 
 of small $r$, which is now plotted in a logarithmic scale.}
\label{fig:final-2}
\end{figure}
In Figs.~\ref{fig:final-1} and \ref{fig:final-2}, we plot the points 
$y = 0.9 y_{\rm max}$ and $0.5 y_{\rm max}$ on the standard $n_s$-$r$ 
plane.  We find that predictions for $n_s$ are consistent with the current 
data, although its detailed values depend on $N_e$ and $p$.  On the other 
hand, the prediction for $r$ depends rather sensitively on $y$, i.e.\ $F$, 
and $p$.  For $N_e = 60$, for example, we obtain
\begin{equation}
  r \gtrsim 
  \left\{ \begin{array}{ll}
    3.8 \times 10^{-3} & (p = 1),\\
    1.1 \times 10^{-4} & (p = 3),\\
    2.7 \times 10^{-6} & (p = 6),
  \end{array} \right.
\label{eq:r_bound}
\end{equation}
for $y < 0.9 y_{\rm max}$, with
\begin{equation}
  r \simeq 
  \left\{ \begin{array}{ll}
    1.2 \times 10^{-2} & (p = 1),\\
    2.2 \times 10^{-3} & (p = 3),\\
    4.7 \times 10^{-4} & (p = 6),
  \end{array} \right.
\label{eq:r_ave}
\end{equation}
for $y = 0.5 y_{\rm max}$. These values of $r$ are interesting 
because a significant portion of the parameter space can be probed by 
next-generation CMB measurements, which are expected to reach $r \sim 
10^{-3}$.  With an improved measurement of $n_s$, this would allow us 
to probe the model further, especially if the value of $p$ is small.

At this point, the power $p$ is unknown.  However, it may be 
determined/constrained if the lattice gauge theory computation of 
$b_4$ is improved by one order of magnitude or more.  The power $p$ 
is given by the ratio between $\bar{b}_4$ and $\bar{b}_2^2$ as%
\footnote{This ratio was also considered in our previous 
 paper~\cite{Nomura:2017ehb}.  Note that we have changed the notation 
 to better match the lattice gauge theory convention.}
\begin{equation}
  \bar{b}_4 = \frac{2(p+2)}{3(p+1)} \bar{b}_2^2 
  \simeq \frac{p+2}{p+1} \times 3.5 \times 10^{-2}.
\label{eq:p-det}
\end{equation}
For relatively small $p$, this relation may be used to determine its 
value.  It may indeed be possible to test the model through an interplay 
between future cosmological observations and lattice gauge theory 
computations.

In conclusion, we have studied predictions for tensor modes in pure 
natural inflation.  We have found that the tensor-to-scalar ratio $r$ 
is bounded from below under natural assumptions.  This originates from 
the fact that the number of metastable branches for $SU(N)$ gauge theory 
is $N$ (a finite value), and hence the field space for inflaton is 
limited.  The actual bound depends on features of the inflaton potential 
that cannot be computed analytically.  The parameter region of the 
model, however, will be constrained by future improvements of lattice 
gauge theory calculations and future observations of CMB radiation. 
It is interesting that physics of strongly coupled gauge theories has 
direct implications for both theory and observations of early universe 
cosmology.

\begin{acknowledgments}
We would like to thank Ryuichiro Kitano, Taizan Watari, Norikazu Yamada, 
and Kazuya Yonekura for related discussion.  This work was supported in 
part by the WPI Research Center Initiative (MEXT, Japan).  The work of 
Y.N. was supported in part by the National Science Foundation under grants 
PHY-1521446, by MEXT KAKENHI Grant Number 15H05895, and by the Department 
of Energy (DOE), Office of Science, Office of High Energy Physics, under 
contract No.\ DE-AC02-05CH11231.  The work of M.Y. was supported in part 
by MEXT KAKENHI Grant Number 15K17634 and JSPS-NRF Joint Research Project.
\end{acknowledgments}


\end{document}